\documentclass[a4paper,superscriptaddress,twocolumn,showpacs,amsmath,amssymb,floatfix,amstex,longbibliography]{revtex4-1}%
\usepackage{graphicx,graphics,rotating}	
\usepackage{xfrac}
\usepackage{dcolumn}
\usepackage[hidelinks]{hyperref}
\usepackage{mathptmx}
\usepackage{array}
\usepackage{verbatim}
\usepackage{amsmath}
\usepackage{mathrsfs}
\usepackage{xcolor}

\newcommand{\abar}{\bar{\alpha}}

\newcommand{\fnu}{f_{\nu}(p/N)}
\newcommand{\crit}{\textrm{crit}}

\definecolor{mygreen}{rgb}{ .01 , .52 ,  .19}
\definecolor{myred}{rgb}{ .71 , 0 ,  0}
\definecolor{oli}{rgb}{ .99 , .39 ,  .19}
\definecolor{mygray}{rgb}{ .60 , .60 ,  .60}
\definecolor{purple}{rgb}{ .70 , .20 ,  .50}
\definecolor{lblue}{rgb}{ .30 , .50 ,  .90}

\usepackage{soul}
\soulregister\ref7
\begin{document}
\title{Symmetry violation of quantum multifractality: Gaussian fluctuations versus algebraic localization}
\author{A. M. Bilen}
\affiliation{Instituto de Investigaciones F\'isicas de Mar del Plata
(IFIMAR), Facultad de Ciencias Exactas y Naturales, Universidad Nacional de Mar del Plata and CONICET,
Funes 3350, B7602AYL
Mar del Plata, Argentina.}

\author{B.~Georgeot}
\affiliation{%
Laboratoire de Physique Th\'eorique, Universit\'e de Toulouse, CNRS, UPS, France
}
\author{O.~Giraud}
\affiliation{Universit\'e Paris Saclay, CNRS, LPTMS, 91405, Orsay, France}
\author{G.~Lemari\'e}
\affiliation{%
Laboratoire de Physique Th\'eorique, Universit\'e de Toulouse, CNRS, UPS, France
}
\affiliation{MajuLab, CNRS-UCA-SU-NUS-NTU International Joint Research Unit, Singapore}
\affiliation{Centre for Quantum Technologies, National University of Singapore, Singapore}
\author{I.~Garc\'ia-Mata}
\affiliation{Instituto de Investigaciones F\'isicas de Mar del Plata
(IFIMAR), Facultad de Ciencias Exactas y Naturales, Universidad Nacional de Mar del Plata and CONICET,
Funes 3350, B7602AYL
Mar del Plata, Argentina.}

\date{\today}%
\begin{abstract}
Quantum multifractality is a fundamental property of systems such as non-interacting disordered systems at an Anderson transition and many-body systems in Hilbert space. Here we discuss the origin of the presence or absence of a fundamental symmetry related to this property.
The anomalous multifractal dimension $\Delta_q$ is used to characterize the structure of  quantum states in such systems. Although the multifractal symmetry relation \mbox{$\Delta_q=\Delta_{1-q}$} is universally fulfilled in many known systems, recently some important examples have emerged where it does not hold. We show that this is the result of two different mechanisms.
The first one was already known and is related to Gaussian fluctuations well described by random matrix theory. The second one, not previously explored, is related to the presence of an algebraically localized envelope. 
While the effect of Gaussian fluctuations can be removed by coarse graining, the second mechanism is robust to such a procedure. We illustrate the violation of the symmetry due to algebraic localization on two systems of very different nature, a 1D Floquet critical system and a model corresponding to Anderson localization on random graphs. 
\end{abstract}
%
\maketitle


The structure of eigenfunctions is a remarkable aspect of quantum systems with critical behavior,
as exemplified by
Anderson localization in disordered quantum systems \cite{anderson1958absence}. 
In that setting, 
consisting of noninteracting particles in a disordered potential, as either disorder or energy is varied, the single-particle wave functions go from being extended to exponentially localized in real space, whereas for a critical value of the parameters they present scale-invariant fluctuations \cite{evers2008anderson}, a property called multifractality \cite{mandelbrot1974,mandelbrot1982,falconer}. 
Recently, it was found that many-body states exhibit multifractal properties in Hilbert space. This is quite generic for ground states \cite{Atas2012,atas2014calculation, luitz2014universal,lindinger2017multifractality,lindinger2019many,pausch2020chaos,lindinger2019many}, but is also characteristic of highly excited states in the many-body localization regime \cite{mace2019multifractal, PhysRevB.102.014208}, a subject of strong interest due to its implications in quantum statistical mechanics\cite{abanin2019colloquium,alet2018many,abanin2017recent,tikhonov2021anderson}. Multifractality has also interesting potential applications for quantum computing \cite{PhysRevX.10.011017}.

Multifractality of quantum states $\psi$ can be defined through the scaling of the average inverse participation ratios (IPR)
\begin{equation}\label{IPR}
I_q(N)\equiv \sum_{r=1}^N \langle |\psi_r|^{2q} \rangle\sim N^{-\Delta_q+1-q},
\end{equation}
with $N$ the size of the system, $q\in\mathbb{R}$ and $\langle\cdot\rangle$ indicating an average over disorder and eigenstates. 
The so-called anomalous dimensions $\Delta_q$ 
quantify the deviation from an ergodic behavior for which $\Delta_q=0$ since \mbox{$\langle|\psi_r|^{2} \rangle\sim N^{-1}$} for all $r$.
For an exponentially localized state, $\Delta_q=1-q$ for $q>0$ and $\Delta_q=-\infty$ for $q<0$. Aside from these two extremes, a multifractal state possesses scale-invariant fluctuations characterized by a set of anomalous dimensions $\Delta_q$ with a non-trivial dependence on $q$.

An important property of quantum multifractality is the symmetry relation
\begin{equation}\label{sym}
\Delta_q=\Delta_{1-q} \,.
\end{equation}
It was first derived from a more general symmetry for the local density of states for nonlinear $\sigma$-models \cite{mirlin2006exact}. 
It can  also be seen as a consequence of the conformal invariance of random critical points \cite{gruzberg2011,gruzberg2013}. An alternative 
viewpoint connects it with fluctuation relations of the Gallavotti-Cohen type~\cite{monthus2009symmetry}. 
These theoretical results and their numerical verification in very distinct physical systems (like
Anderson transitions in 2D and 3D \cite{Mildenberger2007,Obuse2007,Vasquez2008,Rodriguez2008}, quantum Hall transition \cite{Evers2008}, random graphs \cite{deluca2014anderson,tikhonov2016fractality}, and random matrix models \cite{mirlin2006exact,BogGir12,kravtsov2015random})
, with the addition of its experimental verification for the Anderson transition in sound waves \cite{Faez2009}, support the argument in favor of the universality of Eq.~\eqref{sym}. 

Multifractal states are often described as extended but not ergodic, with strong scale-invariant fluctuations. However, in certain systems an overall structure may be present. Exponential localization in finite dimension rules out the scale invariance associated with multifractality. On the other hand, \mbox{\textit{algebraic}} localization, with an envelope characterized by power-law tails, can be compatible with a multifractal behavior. Exponential localization in \textit{infinite} dimension can also be compatible with scale invariance, since the number of sites increases exponentially with distance and compensates the exponential decrease of the wave function.   We will show that the existence of such features does occur in important cases and has physical consequences, in particular for the symmetry \eqref{sym}.  

In this letter we study two important models of a very different nature, whose multifractality violates the symmetry relation Eq.~\eqref{sym} through one and the same mechanism. On the one hand, we focus on a 1D Floquet critical system \cite{BogGir12,MGG,MGGG} arising from the quantization of a classical system with pseudo-integrable dynamics \cite{richens1981} characterized by properties intermediate between chaotic and integrable. The eigenstates, in particular, are multifractal; however, the structure of the states also bears traces of the classical dynamics making this multifractality \emph{inhomogeneous}, with algebraic localization, and leading to the violation of Eq.~\eqref{sym}.
On the other hand, we consider the problem of Anderson localization on random graphs \cite{tikhonov2016fractality,Sonner2017,Sonner2017,tikhonov2016fractality,garcia2020two}, which is deeply related to the problem of many-body localization \cite{tikhonov2021anderson}. The localized phase of this system is multifractal (as is the case in the MBL phase \cite{mace2019multifractal, PhysRevB.102.014208}), a consequence of the combination of typical exponential localization and the exponential proliferation of sites away from localization centers. The resulting multifractality, however, violates Eq.~\eqref{sym}.

\begin{figure*}[!t]
\includegraphics[width=1\textwidth]{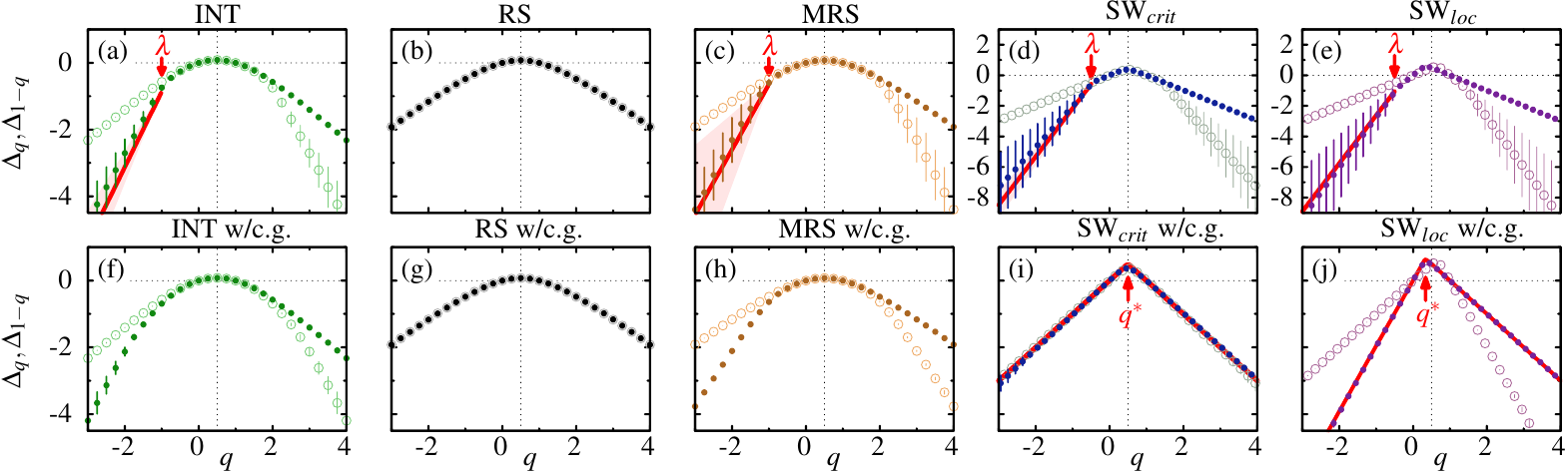}
\caption{$\Delta_q$ (solid circles) and $\Delta_{1-q}$ (empty circles) obtained from the scaling of the average moments $\sum_{p}\left\langle\left|\psi(p)^{2 q}\right|\right\rangle \sim N^{1-q-\Delta_{q}}$
without coarse graining (top row) and with coarse graining (bottom row). The coarse graining parameter is $\ell=16$ for all but the SW model ($\ell=8$).
The model parameters are 
$\gamma=1/3$ (INT), $g=11/24$ (RS), $g=11/24$, $\nu=1$ (MRS), $W=16.75$, $p=0.49$ (SW$_{\crit}$) and $W=21$, $p=0.49$ (SW$_{loc}$). For the INT, RS and MRS models we use all of the $N$ eigenvectors and consider $N=2^{10},...,2^{14}$, and $2^{17}/N$ random realizations; for the SW model $N=2^{10},...,2^{20}$ and the number of realizations ranges from 4000 for the largest $N$ to 27000 for the smallest (using 16 eigenvectors from each realization with energies around the middle of the band $E=0$ obtained using a highly optimized Jacobi-Davidson routine \cite{JacobiDavidson}).
The red arrow (top row) marks the value of $\lambda$ obtained from $P(\alpha)\sim N^{\lambda\alpha}$ for $\alpha\gtrsim 2$ [see Fig.~\ref{fig:palpha}]. The straight red (solid) lines (top row) shows the behavior for $q\lesssim \lambda$ predicted by Eq.~\eqref{eq:nonanalyt} for large $\alpha$. In (i) and (j), for the SW model, the arrows point to $q^*$ which characterizes algebraic localization  and the red (solid) lines show the predicted behavior $\Delta_q=q(1/q^*-1)$ for $q<q^*$ and  $\Delta_q=1-q$ for $q>q^*$. In the critical case (i) $q^*=1/2$ and the symmetry holds.
}
\label{fig:deltaq}
\end{figure*}

We uncover the two main mechanisms behind this effect. The first one, well-known in the literature, is due to random noise described by random matrix theory (RMT) and is commonly discarded as spurious, usually through some form of coarse graining. The second one consists of algebraic localization. It is robust against coarse graining and appears as a characteristic signature of models strongly violating the multifractal symmetry. Remarkably, algebraic localization is able to account for the violation of the symmetry for both the pseudo-integrable model and the localized phase of the Anderson model on random graphs.

The first model we consider is a 1D system periodically kicked by a discontinuous sawtooth potential whose Hamiltonian is  \mbox{$H= P^2 - \gamma \{X\} \sum_{n\in\mathbb{Z}} \delta(t-n)$},
where $\{X\}$ is the fractional part of $X$, and $\gamma$ is the kick amplitude. The stroboscopic dynamics at each period is given by the map \mbox{$P_{n+1}= P_n+\gamma\, ({\rm mod}\,1)$}, \mbox{$X_{n+1}= X_n+2 P_{n+1}\, ({\rm mod}\,1)$}.
For  irrational $\gamma$ the system is ergodic \cite{furstenberg1961strict} while for rational $\gamma\equiv a/b$  (with $a,b$ coprime integers) the classical map is pseudointegrable \cite{richens1981}. 
In the latter case, the motion is periodic in momentum, with $P$ taking only $b$ different values.

The corresponding quantum map is \cite{marklof2000quantum, giraud2004} 
\begin{equation}\label{qmap}
U_{pp'} = \frac{e^{i\phi_p}}{N}\frac{1-e^{i 2 \pi \gamma/h}}{1-e^{2i\pi (p'-p+\gamma/h)/N}} \, ,
\end{equation}
with $p=P/h\in\{0,...,N-1\}$, $\phi_p=-2\pi h p^2$, and $h$ the effective Planck constant $h \equiv 1/N$. The semiclassical limit of this model is $N \rightarrow \infty$ ($h\to 0$). 
In order to get ensemble averages, we replace the phases $\phi_p$ by independent random variables uniformly distributed between 0 and $2\pi$ \cite{bogomolny}. The level spacing statistics of this map is intermediate between Poisson and Wigner-Dyson: it has thus been dubbed the intermediate (INT) map.
In correspondence with the classical dynamics, for $\gamma$ irrational the eigenstates are extended 
\cite{marklof2000quantum, MGGG,BogGir12}, while for rational $\gamma=a/b$ and $N/b\notin\mathbb{Z}$ the eigenstates are multifractal \cite{MGG, MGGG, BogGir12,dubertrand2014two,dubertrand2015multifractality,bilen2019multifractality}. If we replace $h$ by a finite number independent of $N$, \eqref{qmap} corresponds to the Ruijsenaars-Schneider (RS) ensemble of random matrices~\cite{bogomolny2009random} parametrized by $g\equiv \gamma/h$. For the latter model, eigenstates are localized for \mbox{$g=0$}, extended for $g\in\mathbb{Z}\,\backslash\{0\}$, and multifractal otherwise \cite{BogGir12,MGGG12,BogGirPRL}. Varying
$g\in(0,1)$ drives the system from strong to weak multifractality.
The structure of the eigenstates of the INT map exhibits  
remnants of the classical dynamics (seen as $b$ peaks in momentum) intertwined with multifractal fluctuations. In the RS ensemble, however, such a structure is not observed and multifractality is spatially homogeneous. Here we will show that the inhomogeneous multifractal properties present in the INT map are  deeply connected with the violation of \eqref{sym}. The discrepancy in the behavior of these two closely related models will be analysed through a third model defined more precisely later below, the modulated RS (MRS) model, where eigenstates of the RS model are multiplied by an algebraically localized envelope.

Another system showing inhomogeneous multifractal properties corresponds to Anderson localization on random graphs \cite{garcia2017scaling,garcia2020two}. It has generated great interest recently due to its analogy to the many-body localization problem \cite{abou1973selfconsistent,castellani1986upper,mirlin1994distribution,monthus2008anderson,biroli2012difference,deluca2014anderson,monthus2011anderson,facoetti2016non,Sonner2017,parisi2019anderson,kravtsov2018non,savitz2019anderson,tikhonov2016anderson,biroli2018delocalization,garcia2017scaling,tikhonov2019statistics,tikhonov2019critical,kravtsov2015random,altshuler2016nonergodic,bera2018return,biroli2017delocalized,metz2014finite,tikhonov2021anderson}. Here, we consider an Anderson model on a small-world network (SW) \cite{sw1998}, defined (for a fixed parameter $p$) by the Hamiltonian \cite{sw2000,sw2001,sw2005,campo}
\begin{equation}\label{smallworld}
\hat{H} = \sum_{i=1}^N \left(\epsilon_i |i \rangle\langle i| + |i \rangle\langle i+1|\right) + \sum_{k=1}^{\lfloor pN \rfloor} |i_k \rangle\langle j_k| + H.c. \, ,
\end{equation}
corresponding to a chain of $N$ sites (with periodic boundary conditions) with additional long-range links between $\lfloor pN  \rfloor$ randomly chosen pairs of sites $(i_k,j_k)$ with $|i_k-j_k|>1$. The random on-site energies $\epsilon_i$ are sampled from a uniform distribution over $[-W/2,W/2]$. The graph topology of the model is locally treelike, with an average branching number $K\approx 1+2p$, and  when $p\rightarrow1/2$ it approaches a random regular graph. This model has recently been used to study  Anderson localization on random graphs \cite{garcia2017scaling,garcia2020two}. For $W$ larger than some $W_c(p)$ the system is localized, and presents a delocalization transition at $W=W_c(p)$. 
The localized phase has strong multifractality properties \footnote{The localization properties here are distinct from what happens in finite dimensionality where localization implies, in particular, an \emph{exponential} decay of the wave function.}
and the symmetry \eqref{sym} is known not to hold \cite{tikhonov2016fractality, garcia2020two}. 

Figure~\ref{fig:deltaq} shows  $ \Delta_q $ (full circles) and $ \Delta_ {1-q} $ (empty circles) for these systems. Without coarse graining (top row) the symmetry is generally not valid for $ \vert q-1/2 \vert $ large enough. This is true for other systems as well, such as the power-law random banded matrix (PRBM) model \cite{mirlin2006exact}. The only exception we observe is for the RS map, for which Eq.~\eqref{sym} holds perfectly well. The bottom row of Fig.~\ref{fig:deltaq} shows $\Delta_q$ when the wave functions are coarse grained over a small number of sites $\ell$. In this case, $\Delta_q$ is extracted from the scaling of moments $\widetilde{I}_q=\sum_k \mu_k^q$, where $\mu_k=\sum_{r=0}^{\ell-1}|\psi_{k\ell+r}|^2 $ gives the coarse-grained wave-function components. We observe that the coarse graining only restores the symmetry for the SW model at the critical point (SW$_{\crit}$), whereas for the INT map and the SW in the localized regime (SW$_{loc}$) the violation of \eqref{sym} persists.

\begin{figure}[!t]
\includegraphics[width=0.45\textwidth]{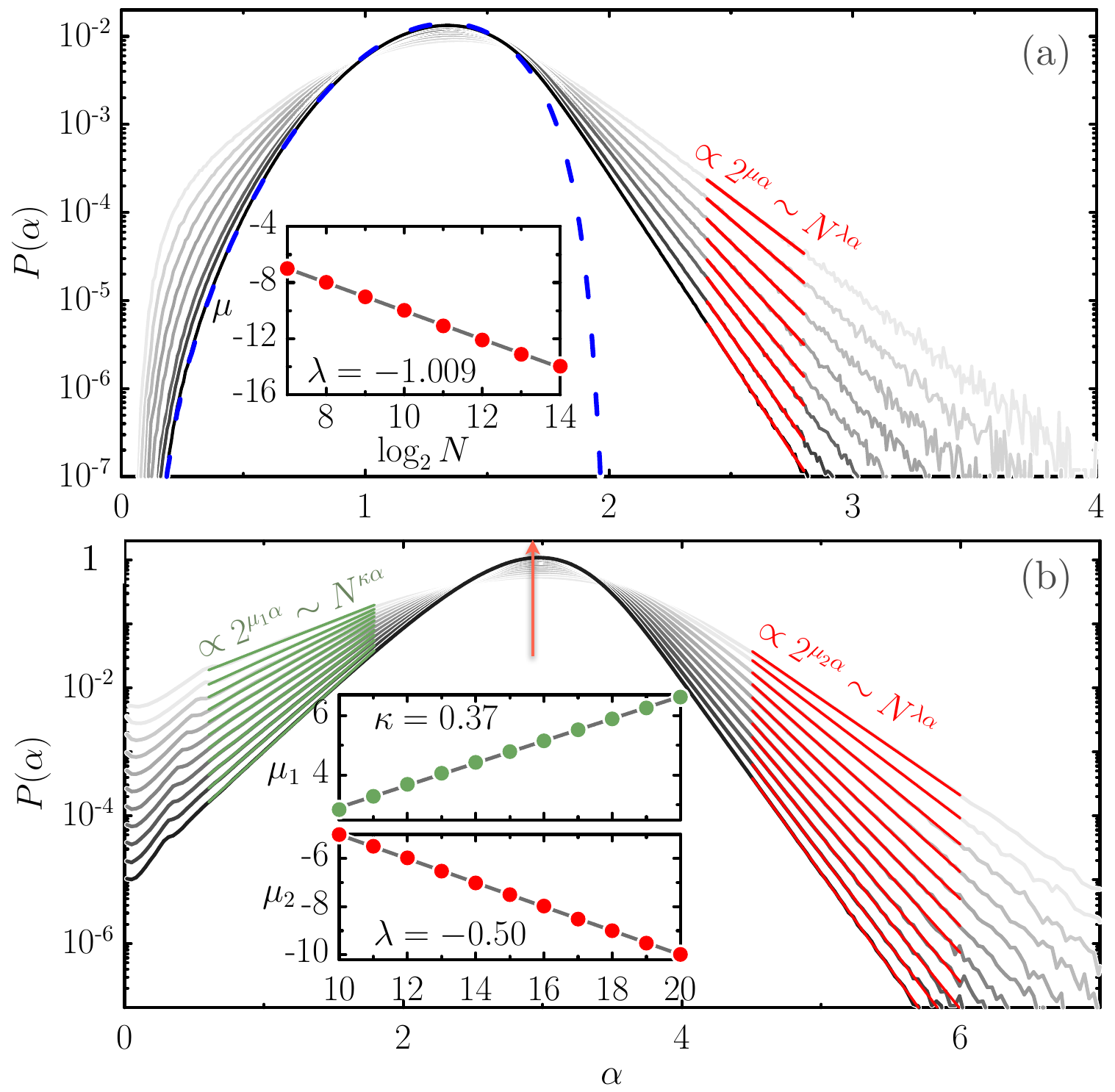} 
\caption{Distribution $P(\alpha)$ for the rescaled eigenfunction amplitudes $\alpha=-\ln |\psi|^2/\ln N$. (a) RS (blue dashed line) for $g=11/24$ and $N=2^{14}$, and MRS model (gray shaded lines) for $g=11/24$, $\nu=1$, and $N=2^{7},...,2^{14}$ (darker shade corresponds to larger $N$). 
The red solid lines show the exponential fit $P(\alpha)\propto 2^{\mu\alpha}$ for large $\alpha$ characterizing the presence of anomalously small values of $|\psi|^2$, due in this case to algebraic localization of type PL1.
The inset shows the exponent $\mu$ vs. $\log_2 N$. A well-defined slope $\lambda$ implies $P(\alpha)\sim N^{\lambda \alpha}$; as can be seen, $\lambda\approx-1=-1/\nu$ as predicted. (b) SW$_{loc}$ with $W=21$, $p=0.49$ and $N=2^{10},...,2^{20}$. Algebraic localization of type PL2 with exponent $-\nu=-1/q^*$ manifests itself in the position of the maximum of $P(\alpha)$ expected to be approximately at $\nu$ (red arrow indicates $\nu$ using the value $q^*\approx0.34 $ extracted from $\Delta_q$; see Fig.~\ref{fig:deltaq}) and the behavior $P(\alpha)\sim N^{\kappa \alpha}$ to the \emph{left} of the maximum (green curves); as shown in the top inset, we find $\kappa=0.37$, which is close to its predicted value $1/\nu=q^*=0.34$. The behavior $P(\alpha)\sim N^{\lambda \alpha}$ to the \emph{right} of the maximum (red curves) is due to Gaussian fluctuations (with $\beta=1$) expected to give $\lambda=-\beta/2=0.50$ in accordance with what we find (bottom inset).}
\label{fig:palpha}
\end{figure}

For the INT and SW$_{crit}$ models without coarse graining, the discontinuity in the derivative of $\Delta_q$ at $q<0$ (indicated with red arrows in Fig.~\ref{fig:deltaq}) signals that the discrepancy comes from the smallest eigenfunction components. To verify this, we study the probability density function $P(\alpha)$ of \mbox{$\alpha\equiv-\ln |\psi|^2/\ln N$}, a quantity that has been used for the study of critical wave function properties at the 3D Anderson transition \cite{rodriguez2009multifractal,Rodriguez2010,rodriguez2011multifractal}.
In Fig.~\ref{fig:palpha}(a) we show  $P(\alpha)$ for two of our models. For the RS model (which is the only one that is symmetric with and without coarse graining) $P(\alpha)$ (blue dashed line) vanishes for $\alpha>2$. By contrast, for the MRS model (defined in detail below) $P(\alpha)$ shows exponential tails \mbox{$P(\alpha)\sim N^{\lambda \alpha}$} for \mbox{$\alpha\gtrsim2$}. We found such tails, with $\lambda<0$ a  system-dependent parameter and for large enough $\alpha$ to the right of the maximum of $P(\alpha)$, for all the other models \cite{papierlong}; they signify the existence of anomalously small wave function amplitudes.

This feature is linked to the singular behavior of $\Delta_q$ at $q<0$. Indeed, assuming
\mbox{$P(\alpha)\sim N^{\lambda \alpha+b}$} over a range \mbox{$\bar{\alpha}<\alpha<\alpha_{\rm max}$} one can show \cite{papierlong} that $\Delta_q$ for $q\lesssim\lambda$ is given by
\begin{equation}
\label{eq:nonanalyt}
\Delta_{q}=(q-\lambda) \alpha_{\max }-q-b, \qquad q \lesssim \lambda\,.
\end{equation}
In particular, this means that the $\Delta_q$ are affected asymmetrically (i.e., ~only for $q\lesssim\lambda$) by the exponential tail of $P(\alpha)$ at large $\alpha$, and hence one can expect that \eqref{sym} will in general not hold (see Appendix). 
This is visible in Fig.~\ref{fig:deltaq}, where red curves are \eqref{eq:nonanalyt} with the parameters extracted from the corresponding $P(\alpha)$.

We now turn to the two mechanisms leading to exponential tails in $P(\alpha)$ for the SW$_{\crit}$ and INT models: RMT Gaussian fluctuations pervading the smallest wave function components, and algebraic localization of eigenfunctions.
The former is usually discarded through coarse graining \footnote{See \cite{deluca2014anderson} for an alternative procedure to get rid of these fluctuations.}. However, the latter  was not known, and is robust against coarse graining.

In the case where the smallest wave function components follow RMT \cite{deluca2014anderson}, their distribution is given by the Porter-Thomas law \cite{porterthomas}, which implies \mbox{$P(\alpha)\propto N^{\beta(1-\alpha)/2}\exp(-\frac12\beta N^{1-\alpha})$}, where $\beta=1,2$ is the Dyson index corresponding to systems with or without time-reversal symmetry, respectively \cite{mehta}. 
For $\alpha>1$ and $N\gg1$, $\ln P(\alpha)$ is thus linear in $\alpha$ with a slope $-\frac12\beta\ln N$, i.e. the type of behavior seen in Fig.~\ref{fig:palpha}(a) with $\lambda$ given by $-\beta/2$.
This is the situation for the SW$_{\crit}$ system, whose $P(\alpha)$ presents an exponential tail $P(\alpha)\sim N^{\lambda \alpha}$ with $\lambda=-1/2$. Time-reversal invariance of \eqref{smallworld} indeed implies that Gaussian fluctuations are of GOE type, i.e.~$\beta=1$. 
As a consequence of Eq.~\eqref{eq:nonanalyt}, $\Delta_q$ displays a discontinuous change in steepness at $q\approx \lambda$, as can be seen in Fig.~\ref{fig:deltaq}(d). The symmetry is recovered after coarse graining [Fig.~\ref{fig:deltaq}(i)], in agreement with the fact that the Gaussian fluctuations are uncorrelated over sites.
We note that in \cite{tikhonov2016fractality} a similar behavior is observed for $P(\alpha)$ at large $\alpha$ for the Cayley tree at the root (with $\beta=2$ due to broken time-reversal invariance). Discarding that part of $P(\alpha)$, as is done in \cite{tikhonov2016fractality}, leads to a symmetric $\Delta_q$ at the transition. Nevertheless, by taking into account the exponential tails at large $\alpha$, a behavior like the one shown in Fig.~\ref{fig:deltaq}(d) for $\Delta_q$ without coarse graining is found for the Cayley tree \cite{papierlong}.

Let us now consider the second mechanism: algebraic localization.
We define algebraic localization as a power-law decay of the wave function characterized by an exponent $\nu>0$. More precisely, we consider the two following types of behavior: (PL1) $|\psi(r)|^2\sim|r-r_0|^{\nu}$ for $|r-r_0|\ll N$ and some fixed point $r_0$; (PL2) $|\psi(r)|^2\sim r^{-\nu}$ with $\nu>0$ for $r\gg 1$. In the first case, the wave function decays to zero around a fixed point of the support with the power-law exponent being positive. In the second case, the exponent is negative, and the decay is asymptotic. In both cases, there exists a certain range of $q$ for which the moments of the wave function will behave in a scale-invariant way, which is why  we can distinguish this type of behavior from that of an exponentially localized state (where the moments either diverge or are system-size independent), and classify it as a special type of multifractality \cite{papierlong}.

The PL1 type of algebraic localization allows us
to better understand the interplay between the classical and the multifractal structure leading to the violation of Eq.~\eqref{sym} for the INT model. First, we note that exponential tails \mbox{$P(\alpha)\sim N^{\lambda \alpha}$} can arise from algebraically localized wave functions of type PL1. Indeed, consider a toy wave function $\phi$ with \mbox{$|\phi(p)|^2=A\left|\frac{p}{N}-P_0\right|^{\nu}$} for $\left|\frac{p}{N}-P_0\right|\ll1$, with $\nu>0$. One can straightforwardly relate the distribution $F(u)$ of small \mbox{$u(p)\equiv|\phi(p)|^2$} to the exponent $\nu$: if \mbox{$p_{\pm}(u)=N\left[\pm(u/A)^{1/\nu}+P_0\right]>0$} is the inverse of $u(p)$ around $P_0$, then $F(u)\approx N^{-1} (p'_{+}(u)+p'_{-}(u))\sim u^{1/\nu-1}$, and thus \mbox{$P(\alpha)\sim N^{-\alpha/\nu}$} giving the exponential tails with $\lambda=-1/\nu$.
With this in mind, we construct a new ensemble of wave functions, which we call the MRS model, by combining RS eigenstates $\psi(p)$ with a smooth PL1-type envelope $\fnu$
to obtain a vector \mbox{$|\widetilde{\psi}(p)|^2=A |\psi(p)|^{2} \fnu$}, 
where $A$ is the normalization.  
We choose \mbox{$\fnu=\left|\sin\left[\pi(\frac{p}{N}-P_0)\right]\right|^{\nu}$}, with $\nu=1$ and $P_0\in (0,1)$ drawn at random for each associated RS eigenstate. 
In Figs.~\ref{fig:deltaq}(c) and \ref{fig:deltaq}(h) we show $\Delta_q$ for the MRS states. As expected from the above discussion, the symmetry is no longer verified. Moreover, $P(\alpha)$, displayed in Fig.~\ref{fig:palpha}(a) (solid grayscale lines), presents exponential tails $P(\alpha)\sim N^{-\alpha/\nu}$. As a consequence, Eq.~\eqref{eq:nonanalyt} applies [see Fig.~\ref{fig:deltaq}(c)]. The modulation of the RS eigenstates with $f_{\nu}$ thus leads to anomalously small wave function components not previously present.

For the SW$_{loc}$ system [i.e.~\eqref{smallworld} in the localized phase], the symmetry is also absent with and without coarse graining [\mbox{Fig.~\ref{fig:deltaq} (e)-(j)}]. As observed from Fig.~\ref{fig:deltaq}(j), it is due to the fact that the maximum of $\Delta_q$ has shifted away from $q=1/2$, thereby precluding $\Delta_q=\Delta_{1-q}$. In this case, the origin of this behavior is an \emph{effective} algebraic localization of type PL2, resulting from the interplay between the typical exponential localization of wave functions $e^{-r/\xi_\text{typ}}$ around their localization center (with the localization length $\xi_\text{typ}$ being the same in almost all directions) and the exponential proliferation $K^r$ of sites at distance $r$~\cite{garcia2020two}. Setting \mbox{$R\equiv K^r$} and \mbox{$q^*=\xi_\text{typ} \ln K$}, we have $e^{-r/\xi_\text{typ}}=R^{-1/q^*}$. The moments for such wave functions can then be rewritten as \mbox{$I_q\sim \int_1^{N/K} dR\, R^{-q/q*}$}, and thus they behave in the same way as for 
PL2-type algebraically localized wave functions $|\psi_R|^2\sim R^{-1/q^*}$ (note that $q^*>0$) of a one-dimensional system of size $\sim N$, known to arise, e.g., for eigenstates of the supercritical PRBM model \cite{yeung1987conjecture,mirlin1996prbm}. The corresponding multifractal spectrum is \mbox{$\Delta_q=q(1/q^*-1)$} for $q<q^*$ and $\Delta_q=1-q$ for $q>q^*$, where $q^*<1/2$ and $q^*\rightarrow1/2$ as the transition is approached \cite{tikhonov2016fractality, garcia2020two}. This behavior is verified in Fig.~\ref{fig:deltaq}(e)-(j). As in the PL1 case, the singular point $q^*$ is related to the exponent $\nu$ of the effective algebraic localization through $q^*=1/\nu$. At the level of $P(\alpha)$, shown in Fig.~\ref{fig:palpha}(b), the PL2-type algebraic localization manifests itself in the position of the maximum of $P(\alpha)$, which (for $\nu>1$) should be approximately located at $\nu=1/q^*$ (red arrow), and in the exponential behavior $P(\alpha)\sim N^{\kappa \alpha}$ with $\kappa=1/\nu=q^*$ (green curves). However, the range of $\alpha$ over which such a behavior holds now stands to the \emph{left} of the maximum of $P(\alpha)$, in contrast to what happens for PL1 and Gaussians fluctuations. Note that this model also presents Gaussian fluctuations, leading to $P(\alpha)\sim N^{\lambda \alpha}$ with $\lambda=-\beta/2$ (red curves) to the \emph{right} of the maximum, as previously discussed.
Finally, we note that the effective algebraic localization of type PL2 that describes the localized phase of the SW model also serves to describe the intermediate non-ergodic fractal phase of the Cayley tree at the root \cite{tikhonov2016fractality}\footnote{It is important to stress that the moments considered in Ref.~\cite {tikhonov2016fractality} are those of the wave function at the \protect \emph  {root}, whereas here we consider the wave function components all around the localization center.}. This is seen by considering that the extension of the exponentially localized behavior no longer scales as the full volume of the system $\sim N$ but scales instead as $\sim N^{\gamma}$ with $0<\gamma<1$ \cite{papierlong}. The resulting $\Delta_q$ is given by $\Delta_q=q\gamma(1/q^*-1)$ for $q<q^*$ and $\Delta_q=(1-q)\gamma$ for $q>q^*$, which coincides with Eq.~(32) of Ref.~\cite{tikhonov2016fractality} with the identification $\gamma=1-\alpha_*$.

The effective algebraic localization in the SW system is reflected in the typical correlation function $ C_\text{typ}(r)=\exp \langle \ln (\sum_{i} \vert \psi_i\vert^2 \vert\psi_{i+r}\vert^2)\rangle$ defined along the 1D lattice. As shown in \cite{garcia2020two}, it decays as $C_\text{typ}(r) \sim \exp(-r/\xi_\text{typ})=R^{-1/q^*}$, and therefore algebraically with the number $R=K^r$ of pairs of sites at a distance $r$ of a given site.

\begin{figure}[!t]
\includegraphics[width=0.9\linewidth]{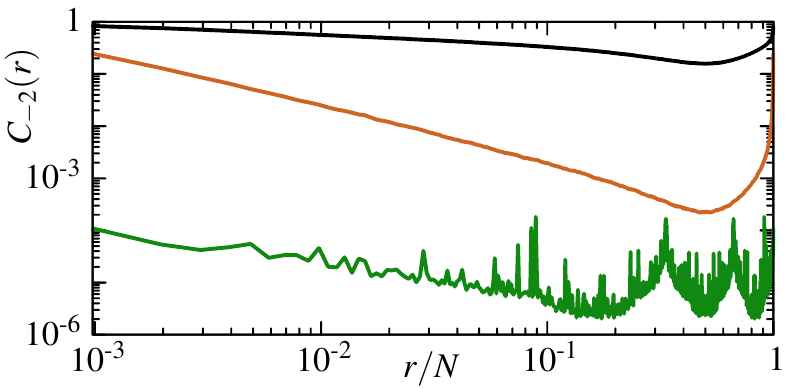} 
\caption{Average correlation function $C_{-2}(r)$ in the RS (top, black), MRS (middle, brown), and INT models (bottom, green). System size is $N=2^{10}$ and number of random realizations $128$. System parameters are $g=11/24$, $\nu=1$ and $\gamma=1/3$. Strong correlation in the RS model is associated with the cutoff in $P(\alpha)$ that favors the symmetry. Algebraic localization in MRS translates into less correlated eigenfunctions and violation of the symmetry. The INT map has both algebraic localization and Gaussian fluctuations.}
\label{fig:corr}
\end{figure}

The difference between
Gaussian fluctuations and algebraic localization in the INT and RS models can also be visualized considering the average correlation function   \mbox{$C_q(r)=(N\langle|\psi|^{2q}\rangle)^{-1}\sum_{r'}\langle |\psi_{r'}|^q |\psi_{r'+r}|^q\rangle $}. As we are interested in the smallest wave function components, in Fig.~\ref{fig:corr}
we consider the case $q<0$. Gaussian fluctuations are spatially uncorrelated and are therefore associated with a very low constant value of $C_{q<0}(r)$ for $r>0$. On the contrary, algebraic localization is associated with a power-law decay of the correlation function which enhances the usual algebraic decrease due to multifractality \cite{Evers2008}. In  
Fig.~\ref{fig:corr}, the MRS model (brown line) shows a faster algebraic decay than the RS model (black line), an effect of the envelope $f_{\nu}$.
The INT model (green line) has a similar decay as the MRS model, but multiplied by a very small constant prefactor, a signature of the presence of both  algebraic localization and Gaussian fluctuations. In contrast, Gaussian fluctuations are absent in the RS model, consistent with the symmetry \eqref{sym} holding without coarse graining.

In summary, we have studied the multifractal symmetry $\Delta_q=\Delta_{1-q}$ known to be valid for a large class of systems at criticality, and found two mechanisms that lead to its violation. 
The first one concerns anomalously small wave function components that are due to Gaussian fluctuations at small scales. Though such fluctuations are known and eliminated by coarse graining, we showed that their precise characteristics can be obtained through their effect on multifractality.  The second mechanism corresponds to algebraic localization of an otherwise homogeneously multifractal wave function that generates anomalously small values of $|\psi|^2$. In that case, the symmetry violation is robust to coarse graining.

We find that the robust violation of 
the multifractal symmetry is intrinsically related to \textit{inhomogeneous} multifractal properties. 
For the Floquet critical system in the semiclassical regime our results suggest that the wave functions are modulated by an envelope with algebraic behavior around local minima, inducing a spatial inhomogeneity of the multifractal fluctuations without destroying their characteristic scale invariance. We interpret this envelope as similar to EBK envelopes \cite{berry77}, localized around regular tori in integrable systems. Experimentally, the properties of the Floquet system we have studied could be assessed in photonic crystal implementations \cite{schwartz2007transport,lahini2008anderson,levi2012hyper} or in cold atom experiments \cite{lemarie2010,lopez2013,sagi2012,chabe2008experimental,lemarie2009observation}, both of which have been of paramount importance in the experimental study of Anderson localization. Pseudo-integrable billiards, for which multifractal dimensions have been calculated numerically \cite{BogSch04b}, are also amenable to experiments, for instance via electromagnetic microwaves in cylindrical cavities \cite{Micexp92}, or liquid crystal smectic films \cite{smectic}.

Anderson localization in random graphs is also associated with an effective algebraic localization due to the interplay between the exponential localization of wave functions and the exponential growth of available sites, leading to strong multifractality in the localized phase where the symmetry relation is not respected.
The non-ergodic properties of this problem have been much discussed recently in relation to the many-body localization problem \cite{abou1973selfconsistent,castellani1986upper,mirlin1994distribution,monthus2008anderson,biroli2012difference,deluca2014anderson,monthus2011anderson,facoetti2016non,Sonner2017,parisi2019anderson,kravtsov2018non,tikhonov2016anderson,biroli2018delocalization,garcia2017scaling,tikhonov2019statistics,tikhonov2019critical,kravtsov2015random,altshuler2016nonergodic,bera2018return,biroli2017delocalized}, the Cayley tree \cite{tikhonov2016fractality, Sonner2017, biroli2018delocalization, kravtsov2018non,savitz2019anderson} and certain types of random matrices \cite{Kravtsov2015, facoetti2016non, PhysRevResearch.2.043346, biroli2020levy, bogomolny2020statistical}. Also, the MBL phase \cite{mace2019multifractal, PhysRevB.102.014208} and many-body eigenstates \cite{luitz2014universal,lindinger2017multifractality,lindinger2019many,pausch2020chaos} are generically multifractal in Hilbert space. It would be interesting to study these important problems along the lines of this paper.

\begin{acknowledgments}
 This study has been (partially) supported through the EUR Grant NanoX ANR-17-EURE-0009 in the framework of the "Programme des Investissements d'Avenir", the French-Argentinian LIA LICOQ, and also by research funding Grants No.~ANR-17-CE30-0024, ANR-18-CE30-0017 and ANR-19-CE30-0013. We thank Calcul en Midi-Pyr\'en\'ees (CALMIP) for computational resources and assistance. A.M.B.~is grateful to Nicolas Mac\'e for valuable technical advice concerning the numerical computations, and to Pablo Kaluza for useful discussions. I.G.-M.~received funding from CONICET (Grant No.~PIP 11220150100493CO) and ANCyPT (Grant No.  PICT-2016-1056).
\end{acknowledgments}

\appendix*
\section*{Appendix: $P(\alpha)$ at large $\alpha$ and symmetry of $\Delta_q$} 
In this appendix we discuss in more detail the relationship between the distribution $P(\alpha)$ at large $\alpha$ and the symmetry $\Delta_{q}=\Delta_{1-q}$.

We first observe that the distribution $P(\alpha)$ at large $\alpha$ affects $\Delta_q$ at small $q$. In particular, 
when \mbox{$P(\alpha)\sim N^{\lambda \alpha+b}$} for $\alpha \in [ \bar{\alpha},\alpha_{max}]$, with $\lambda<0$,
we have $\Delta_q=(q-\lambda) \alpha_{\max }-q-b$ for $q \lesssim \lambda < 0$ [see Eq.~(5)]. {Thus, the function $\Delta_q$ in this region $q \lesssim \lambda < 0$ can be expressed in terms of the parameters $\lambda$ and $b$ that govern the exponential tail of $P(\alpha)$ at $\alpha>\bar{\alpha}$}. For $q>0$, on the other hand, the behavior {of $\Delta_q$} is dependent only on the properties of $P(\alpha)$ for $\alpha<\bar{\alpha}$.
As a consequence, given a $\Delta_q$ that satisfies the symmetry $\Delta_q=\Delta_{1-q}$,
any modification of $P(\alpha)$ on one side of $\abar$ only will result in the violation of this symmetry.

Suppose now that we have two systems for which the distributions $P(\alpha)$ coincide for $\alpha<\abar$, but with $P(\alpha)$ vanishing at $\alpha>\abar$ for one of them and $P(\alpha)\sim N^{\lambda \alpha}$ at $\alpha>\abar$ for the other one. {This is illustrated in Fig.~2 of the main text, where the dashed line vanishes at $\abar=2$ while the solid lines display exponential tails beyond that value.}
It follows from the {considerations} above that if one of these systems satisfies the symmetry $\Delta_q=\Delta_{1-q}$ then the other one will not.

An instance of this is {precisely given by the models whose distributions are displayed in Fig.~2: the} Ruijsenaars-Schneider (RS) {random matrix} model and the modulated Ruijsenaars-Schneider (MRS) vector ensemble. As can be seen in Fig.~2 the $P(\alpha)$ at $\alpha\lesssim2$ are almost identical (and all the more so as $N$ gets larger), whereas for $\alpha\gtrsim2$ the distribution vanishes for the RS model while it behaves as $P(\alpha)\sim N^{\lambda \alpha}$ for the MRS model. Since $\Delta_q$ is symmetric for the RS model, this implies the violation of the symmetry for the MRS model. This is confirmed by the plots
in Figs.~1(b) and 1(c).

Another example is the small-world model at criticality (SW$_{\crit}$), where coarse graining only affects the tail of $P(\alpha)$ at $\alpha>\abar$. Since with coarse graining the symmetry holds, in the absence of it the symmetry breaks down [see Fig.~1(d) and 1(i)]. In the RS model, the coarse graining does not affect $P(\alpha)$ for $\alpha>\abar$ and the symmetry remains valid. 

%

\end{document}